\documentclass[twocolumn,showpacs,showkeys,preprintnumbers,superscriptaddress,amsmath,floatfix,amssymb,secnumarabic,nofootinbib]{revtex4}
\usepackage[colorlinks=true]{hyperref}
\usepackage{graphicx}
\usepackage{braket}
\usepackage{amsmath}
\usepackage{epsfig}
\usepackage{float}
\usepackage{caption}
\usepackage{subcaption}

\newcommand{\ev}{\, {\rm eV}}

\def\({\left(}
\def\){\right)}
\def\[{\left[}
\def\]{\right]}

\newcommand{\lr}[1]{ \left( #1 \right) }
\newcommand{\lrs}[1]{ \left[ #1 \right] }

\newcommand{\tr}{ {\rm Tr} \, }

\newcommand{\beq} {\begin{eqnarray}}
\newcommand{\eeq} {\end{eqnarray}}

\newcommand{\comment}[1]{}

\begin{document}
\sloppy

\title{Quantum Monte Carlo study of static potential in graphene}

\author{N.\,Yu.~Astrakhantsev}
\email[]{nikita.astrakhantsev@itep.ru}
\affiliation{Institute of Theoretical and Experimental Physics, 117218 Moscow, Russia}
\affiliation{Moscow Institute of Physics and Technology, Institutskii per. 9, Dolgoprudny, Moscow Region, 141700 Russia}

\author{V.\,V.~Braguta}
\email[]{braguta@itep.ru}
\affiliation{Institute of Theoretical and Experimental Physics, 117218 Moscow, Russia}
\affiliation{Moscow Institute of Physics and Technology, Institutskii per. 9, Dolgoprudny, Moscow Region, 141700 Russia}
\affiliation{Far Eastern Federal University, School of Biomedicine, 690950 Vladivostok, Russia}
\affiliation{Institute for High Energy Physics NRC "Kurchatov Institute", Protvino, 142281 Russian Federation}
\affiliation{Bogoliubov Laboratory of Theoretical Physics, Joint Institute for Nuclear Research, Dubna, Russia}

\author{M.\,I.~Katsnelson}
\affiliation{Radboud University, Institute for Molecules and Materials,
Heyendaalseweg 135, NL-6525AJ Nijmegen, The Netherlands}
\affiliation{Ural Federal University, Theoretical Physics and Applied Mathematics Department, Mira Str. 19,  620002 Ekaterinburg, Russia}

\author{A.\,A.~Nikolaev}
\affiliation{Institute of Theoretical and Experimental Physics, 117218 Moscow, Russia}
\affiliation{Far Eastern Federal University,  School of Biomedicine, 690950 Vladivostok, Russia}

\author{M.\,V.~Ulybyshev}
\affiliation{Institute of Theoretical Physics, University of Regensburg, D-93053 Germany, Regensburg, Universitatsstrasse 31}

\begin{abstract}

In this paper the interaction potential between static charges in suspended graphene is studied within the quantum Monte Carlo approach. We calculated the dielectric permittivity of suspended graphene for the set of temperatures and extrapolated our results to zero temperature. The dielectric permittivity at zero temperature has the following properties. At zero distance $\epsilon=2.24\pm0.02$. Then it rises and at a large distance the dielectric permittivity reaches the plateau $\epsilon\simeq4.20\pm0.66$. The results obtained in this paper  
allow to draw a conclusion that full account of many-body effects in the dielectric 
permittivity of suspended graphene gives $\epsilon$ very close to 
the one-loop results. Contrary to the one-loop result, the two-loop prediction for the dielectric permittivity deviates from our result. So, one can expect 
large higher order corrections to the two-loop prediction for the dielectric permittivity of suspended
graphene.
\end{abstract}
\pacs{73.22.Pr, 05.10.Ln, 11.15.Ha}
\keywords{graphene, electron properties, Coulomb interaction, Monte-Carlo simulations}

\maketitle

\section{Introduction}

Graphene, two dimensional crystal composed of carbon atoms packed in a honeycomb lattice~\cite{Novoselov:04:1, Geim:07:1}, attracts considerable interest due to its electronic properties. 
There are two Fermi points in the electronic spectrum of graphene. In the vicinity of each point the fermion excitations are similar to the massless Dirac fermions living in two dimensions \cite{wallace,semenoff,Novoselov:05:1,zhang}.
Relativistic nature of fermion excitations in graphene leads to numerous quantum relativistic phenomena 
such as Klein tunneling, minimal conductivity through evanescent waves, relativistic collapse at a supercritical 
charge and etc.~\cite{Guinea, Kotov:2, Katsnelson}.

The Fermi velocity in graphene is much smaller than the speed of light ($v_F \sim c/300$), resulting in negligible retardation effects and magnetic interaction between quasiparticles. Thus the 
interaction in graphene can be approximated by the instantaneous Coulomb potential with the large effective coupling 
constant $\alpha_{eff}=\alpha \cdot c/v_F \sim 300/137 \sim 2.2$. 

It is reasonable to assume that the observables in graphene theory are considerably renormalized due to strong interaction as compared to the non-interacting theory. 
For instance, the leading perturbative renormalization
of the Fermi velocity with the logarithmic accuracy~\cite{Gonzalez:1993uz, Mishchenko, Herbut, 
Fritz, Fernando} leads to the increase of $v_F$ as large as $\sim 100 \%$ for suspended graphene.  
One can expect that higher order renormalization leads to further considerable change of the 
bare value of the Fermi velocity. However, existing experimental measurements
of the Fermi velocity~\cite{Elias,PNAS} are in a good agreement with the first-order perturbation theory
improved by the one-loop expression for the dielectric permittivity of graphene. 
Recently it was shown that the next-to-leading order corrections in the random
phase approximation (RPA) are small relative to the leading-order RPA results~\cite{dassarma}.
Nevertheless it is not clear what happens to the perturbative corrections 
after the next-to-leading order. 

Another important observable in graphene is the interaction potential between static charges. 
The renormalization of the static potential can be parameterized by the dielectric permittivity depending on the distance between static charges $\epsilon(r)$. One-loop expression 
for $\epsilon(r)$ was calculated in \cite{Astrakhantsev:2015cla}. 
At small distances $\epsilon(r) \sim 2$. Then the dielectric permittivity grows with the distance and at large distances $r \gg a$ \footnote{The $a$ is the distance between carbon atoms in graphene.} 
it reaches well known one-loop expression \cite{Katsnelson,Kotov:2}
\beq
\epsilon=1 + \frac {\pi} 2 \alpha_{eff}. 
\label{eps_one_loop}
\eeq   
For $\alpha_{eff} = 2.2$ this formula gives $\epsilon = 4.4$. So, one sees 
that the one-loop correction is very large. Two-loop correction to $\epsilon$ 
was calculated in \cite{fogler} and can be written as 
\beq
\epsilon=1 + \frac {\pi} 2 \alpha_{eff} + 0.778 \alpha_{eff}^2.
\label{eps_two_loop}
\eeq
If one takes $\alpha_{eff} = 2.2$ the dielectric permittivity is $\epsilon=8.2$.
If one accounts one-loop renormalization of $\alpha_{eff}$ (see below), $\epsilon\simeq2.5$.
So, it is clear that the next-to-leading order corrections considerably modify the one-loop result. 
One can also expect that higher order corrections are also significant. 

In this paper we are going to study the interaction potential between static charges in suspended graphene 
within quantum Monte Carlo simulations (see \cite{Ulybyshev:2013swa} for details).
The first Monte Carlo study of the static interaction potential based on the low energy effective theory of graphene 
was done in paper \cite{Braguta:2013rna}.  
In this paper we are going to use the approach which is based on the tight-binding Hamiltonian 
without the expansion near the Fermi points. This approach allows to avoid ambiguity due to regularization procedure. 
In addition, the interactions between quasiparticles are parameterized by the realistic phenomenological 
potential obtained in \cite{Wehling} that significantly deviates from the Coulomb potential at small distances. 
However, the main advantage of the Monte-Carlo approach is that it fully accounts interactions between 
quasiparticles basing on the first principles.

This paper is organized as follows. In the next section we briefly describe the method of 
lattice Monte Carlo simulation of graphene. Section III is devoted to the discussion of how the 
potential of static charges can be calculated within Monte Carlo simulations. 
The results of the calculation are presented in Section IV. In the last section we summarize our results.

\label{IntroductionSec}

\section{Brief description of the model}
\label{ModelSec}

In the calculation we use the tight-binding model of graphene. The Hamiltonian consists of the tight-binding term and the interaction part describing the full electrostatic interaction between quasiparticles:
\begin{eqnarray}
\label{tbHam}
 \hat{H}= - \kappa \sum\limits_{<x,y>} \lr{ \hat{a}^{\dag}_{y, \uparrow} \hat{a}_{x, \uparrow}
+  \hat{a}^{\dag}_{y, \downarrow} \hat{a}_{x, \downarrow} + h.c.}
 \nonumber \\ +
 \sum\limits_{x=\{1,\xi\}} m (\hat{a}^{\dag}_{x, \uparrow} \hat{a}_{x, \uparrow} - \hat{a}^{\dag}_{x, \downarrow} \hat{a}_{x, \downarrow} ) \nonumber \\ - \sum\limits_{x=\{2,\xi\}} m (\hat{a}^{\dag}_{x, \uparrow} \hat{a}_{x, \uparrow} - \hat{a}^{\dag}_{x, \downarrow} \hat{a}_{x, \downarrow} ) \nonumber \\ +{1 \over 2} \, \sum\limits_{x,y} V_{xy} \hat{q}_x \hat{q}_y,
\end{eqnarray}
where $\kappa = 2.7 \ev$ is the hopping between nearest-neighbors, $\hat{a}^{\dag}_{x, \uparrow}$, $\hat{a}_{x, \uparrow}$ and $\hat{a}^{\dag}_{x, \downarrow}$, $\hat{a}_{x, \downarrow}$ are creation/annihilation operators for spin-up and spin-down electrons at $\pi$-orbitals. Spatial index $x=\{s, \xi\}$ consists of sublattice index $s=1,2$ and two-dimensional coordinate $\xi=\{\xi_1, \xi_2\}$ of the unit cell in rhombic lattice. Periodic boundary conditions are imposed in both spatial directions in the manner of \cite{Ulybyshev:2013swa}. The mass term has different sign at two sublattices. According to the simulation algorithm, one should introduce the mass in order to eliminate zero modes from the fermionic determinant. The calculation is carried out at few various masses and 
the final results are obtained through the chiral extrapolation $m\to0$.

The matrix $V_{xy}$ is the bare electrostatic interaction potential between sites with coordinates $x$ and $y$ and $\hat{q}_x = \hat{a}^{\dag}_{x, \uparrow} \hat{a}_{x, \uparrow} + \hat{a}^{\dag}_{x, \downarrow} \hat{a}_{x, \downarrow} - 1$ is the electric charge operator at lattice site $x$. 
The potential $V_{xy}$ represents the screened Coulomb interaction. At small distances $r/a \leqslant 2$ we employ phenomenological potentials $V_{00},\,V_{01},\,V_{02},\,V_{03}$ calculated in \cite{Wehling}, while at distances $r/a \geqslant 2$ we use the Coulomb-like potential
\begin{equation}
	\label{eq:potentials_large}
    V(r) = \frac{A}{r / a + C},
\end{equation}
where $A =\alpha \cdot \hbar c/a= 10.14\,$eV, $C = 0.82$. The parameter $C$ is chosen so that $V(2 a) = V_{03}$. This choice ensures smooth interpolation between regions of the phenomenological potential and the Coulomb-like potential. 

All calculations were performed using the Hybrid Monte-Carlo algorithm. Details of the algorithm are described in \cite{Ulybyshev:2013swa}. 
The method is based on the Suzuki-Trotter decomposition. Partition function $\exp (-\beta \hat H)$ is represented in the form of a functional integral in Euclidean time. Inverse temperature is equal to the number of time slices multiplied by the step in Euclidean time: $\delta \tau L_t=\beta=1/T$. Since the algorithm requires fermionic fields to be integrated out, we eliminate all the four-fermionic terms in the full Hamiltonian using the Hubbard-Stratonovich transformation. The final form of the partition function can be written as:
\begin{eqnarray}
\label{PartFunc2}
 \tr e^{-\beta \hat{H}} \cong \int \mathcal{D}\varphi_{x,n} e^{-S\lrs{\varphi_{x,n}}}
 |\det{M\lrs{\varphi_{x,n}}}|^2\,,
\end{eqnarray}
where $\varphi_{x,n}$ is the Hubbard-Stratonovich field for timeslice $n$ and spatial coordinate $x$. Particular form of the fermionic operator $M$ is described in \cite{Ulybyshev:2013swa}. The absence of the sign problem (appearance of the squared modulus of the determinant) is guaranteed by the particle-hole symmetry in graphene at neutrality point. Action for the Hubbard field $S\lrs{\varphi_{x,n}}$ is also a positively defined quadratic form for all choices of the electron-electron interaction used in our paper. Thus we can generate configurations of $\varphi_{x,n}$ by the Monte-Carlo method using the weight (\ref{PartFunc2}) and calculate physical quantities as averages over generated configurations.

\begin{figure}[t]
    \centering
    \includegraphics[scale = 0.55]{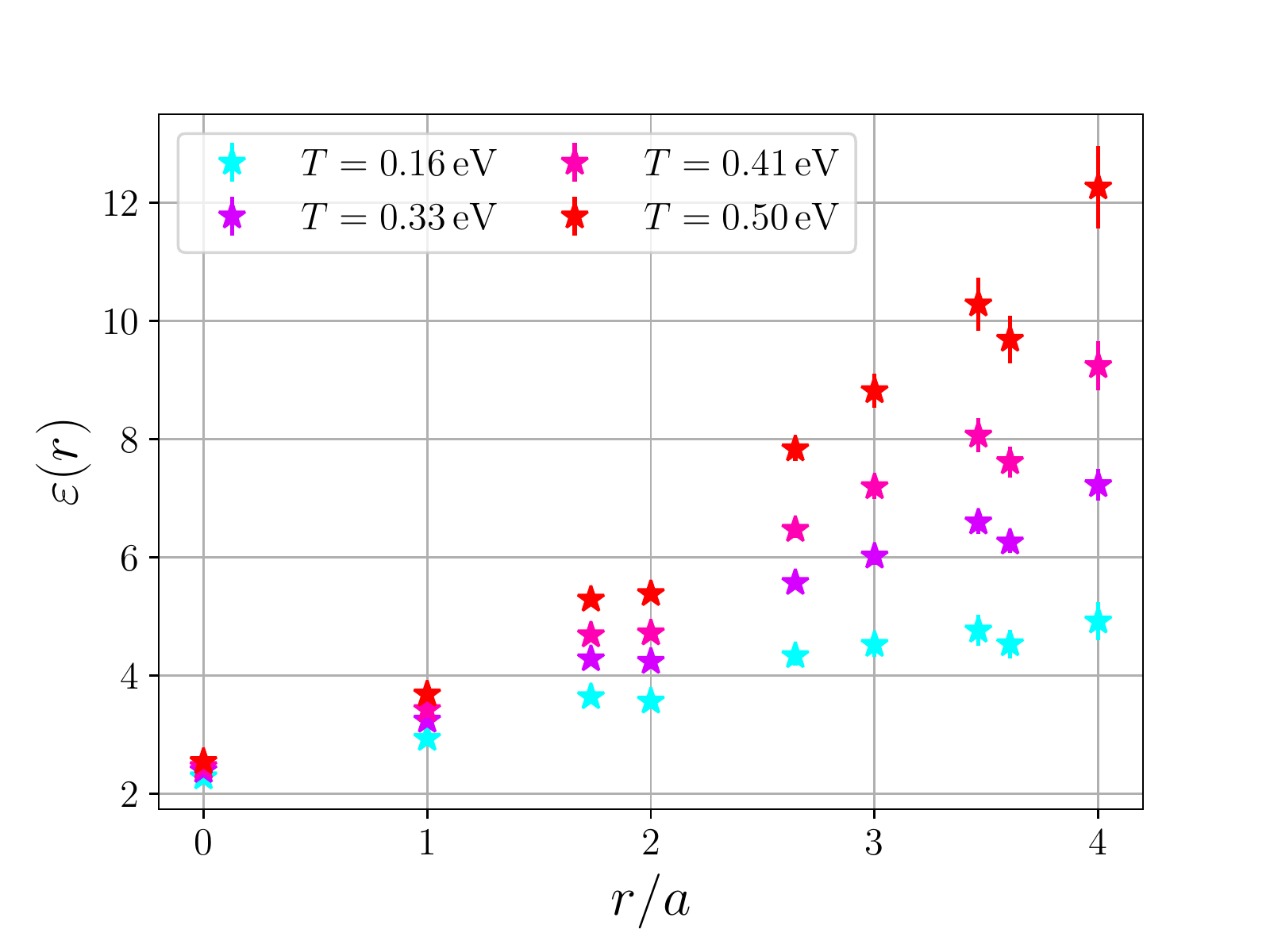}
    \caption{The dielectric permittivity of graphene $\epsilon(r, T)$ as a function of distance $r/a$ for $T = 0.167,\,0.333,\,0.417,\,0.5\,$eV.}
    \label{fig:chiral_potentials}
\end{figure}

\section{Details of the calculation}
\label{DetailsSec}
To calculate the potential of the static charges in graphene one introduces the Polyakov loop of the Hubbard-Stratonovich field. 
The Polyakov loop is defined as a product of the factors\footnote{The charge $Q$ is measured in units of electron charge $e$.}
 $\exp {(i Q \delta \tau \varphi_{\vec{x}, t})}$ over all 
slices in Eucledean time $t$ and with fixed spatial coordinate $\vec{x}$
\begin{equation}
	\label{eq:P_loop_def}
	\displaystyle L_Q(\vec{x}) = \prod_{t = 0}^{L_t - 1} \text{exp}(- i 	Q \Delta \tau \varphi_{\vec{x}, t}).
\end{equation}
Physically the introduction of the operator $\displaystyle L_Q(\vec{x})$ implies the 
calculation of the partition function of graphene with the static charge $Q$
\begin{equation}
	\label{eq:P_loop_self}
	\left\langle L_Q \right\rangle = \text{exp}(- F_Q / T)\,,
\end{equation}
where $F_Q$ is the free energy of the static charge $Q$ in graphene. 

\begin{figure}[t]
        \centering
        \includegraphics[scale = 0.55]{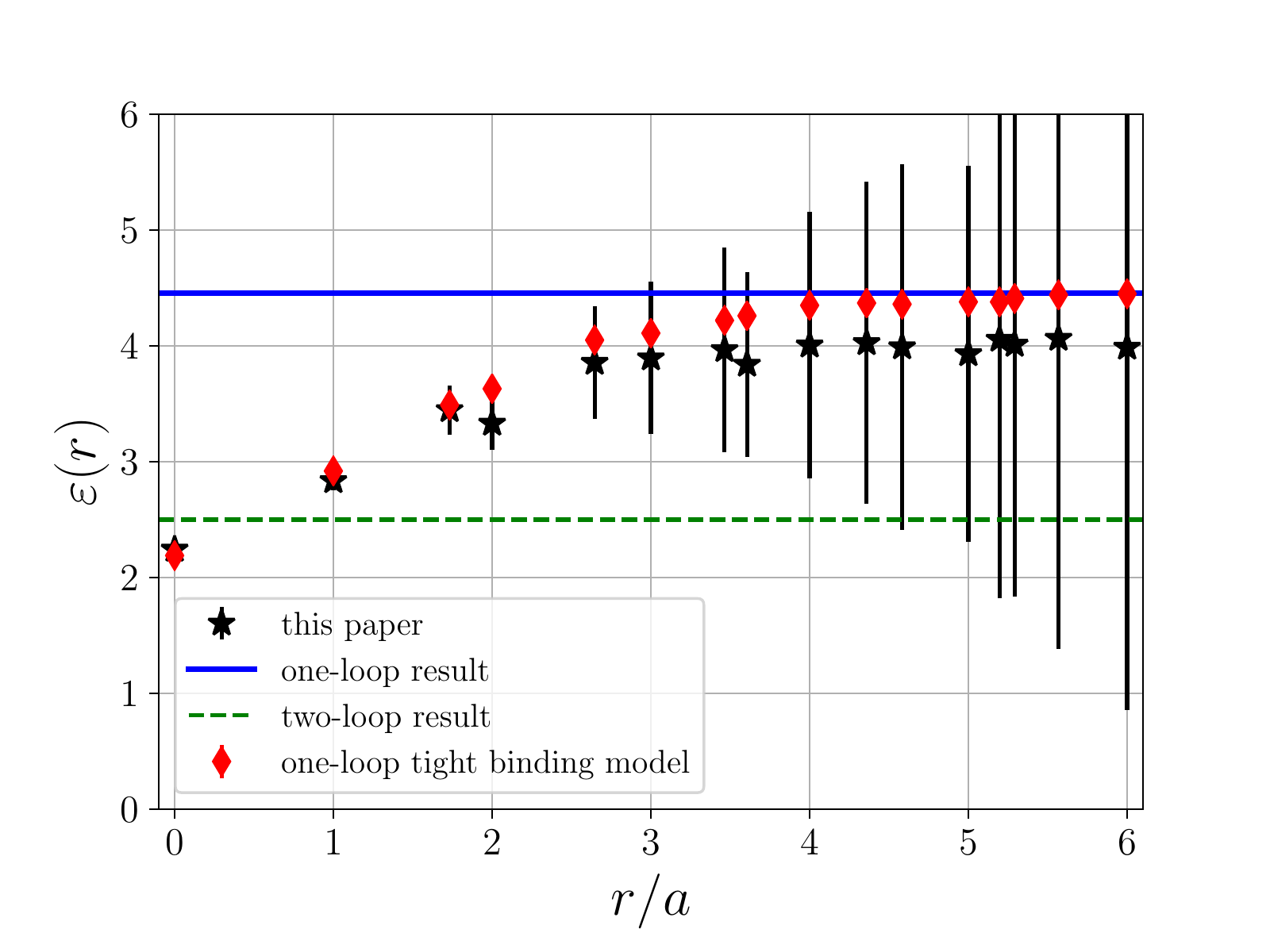}
        \caption{The dielectric permittivity of suspended graphene $\epsilon(r)$ as a function of distance $r/a$ at zero temperature.
The results obtained in this paper are represented by black stars.
The blue line and green dashed line correspond to the one-loop result (\ref{eps_one_loop}) and the two-loops result (\ref{eps_two_loop})
correspondingly. The red diamond points correspond to the one-loop calculation of the $\epsilon(r)$ based on the tight binding 
model (\ref{tbHam}) carried out in \cite{Astrakhantsev:2015cla}.
}
        \label{fig:zero_potentials}
\end{figure}

Similarly the correlation function of the Polyakov loops $\langle L_Q(0) L_Q^*(\vec{r}) \rangle$ is determined by the free 
energy of static charges $Q$ and $-Q$ separated by the distance $\vec{r}$. The free energy of this system is determined by
the potential of the static charges in graphene $V_{QQ}(\vec{r})$. Thus we have
\begin{equation}
	\label{eq:P_loop_corr}
	\left\langle L_Q(0) L_Q^*(\vec{r}) \right\rangle \sim \text{exp}(- V_{QQ} (\vec{r}) / T)\,.
\end{equation}
In order to obtain interaction potential we use the following relation
\begin{equation}
	\label{eq:V_qq_def}
	V_{QQ}(\vec{r}) = - \frac{T}{Q^2}(\text{ln}\left\langle L_Q(0) L_Q^*	(\vec{r}) \right\rangle - 2\text{ln}\left\langle L_Q \right\rangle)\,.
\end{equation}
In the simulations we considered $Q = 0.1$ which significantly improves signal-to-noise ratio both for the correlator and for the Polyakov loop.
This value of $Q$ was taken from \cite{Braguta:2013rna}.

Monte Carlo simulation of graphene was carried out on the lattices with spatial extension $L_x = L_y = 30$. 
The lattice spacing in temporal direction is $\delta \tau=0.1\,$eV$^{-1}$. 
The temporal sizes of the lattices under study are $L_t = 60,\,50,\,36,\,34,\,30,\,26,\,24,\,22,\,20$ which correspond to the temperatures $T = 0.167,\,0.2,\,0.278,\,0.294,\,0.333,\,0.385,\,0.417,\,0.455,\,0.5\,$eV. For the lattices $30^2 \times 36 \ldots 20$ we conducted the calculations 
at the following values of the fermion mass $m = 0.03,\,0.05,\,0.07,\,0.1\,$eV. 
For the two lowest temperatures on the lattices $30^2 \times 60$ and $30^2 \times 50$ 
the fermion masses were $0.01,\,0.02,\,0.03,\,0.04\,$eV. For these values of the fermion 
masses we simulate relativistic fermions. To obtain the results for the massless fermions
we fit our data for all temperatures and distances under study by the function $V_{QQ}(r)=V_0(r)+V_1(r) m^2$.
For all temperatures and distances the data is well described by this fit ($\chi^2 \leq 1$). The function $V_0(r)$ gives the potential 
in the massless limit. Below we present the results obtained in the limit of massless fermions $m\to0$.

\begin{table}[t]
 \begin{tabular}{| c | c | c |}
    \hline
    $r/a$ & $\varepsilon(r)$ & $\varepsilon_{\mbox{1 loop}}(r)$ \\ \hline  
$0.00$ & $2.24 \pm 0.02$ & $2.19$ \\ \hline
$1.00$ & $2.83 \pm 0.08$ & $2.92$ \\ \hline
$1.73$ & $3.45 \pm 0.21$ & $3.49$ \\ \hline
$2.00$ & $3.33 \pm 0.23$ & $3.63$ \\ \hline
$2.65$ & $3.86 \pm 0.49$ & $4.05$ \\ \hline
$3.00$ & $3.89 \pm 0.66$ & $4.11$ \\ \hline
$3.46$ & $3.97 \pm 0.88$ & $4.22$ \\ \hline
$3.61$ & $3.84 \pm 0.80$ & $4.26$ \\ \hline
$4.00$ & $4.01 \pm 1.15$ & $4.35$ \\ \hline
    \end{tabular}
\caption{The dielectric permittivity of suspended graphene $\epsilon(r)$ as a function of distance $r/a$ at zero temperature.
The first column is the distance in graphene lattice units. The second column is the $\epsilon(r)$ calculated in this paper.
The third column contains the $\epsilon(r)$ calculated at the one-loop level within the tight binding model (\ref{tbHam}).}
        \label{tab:zero_potentials}
\end{table}

\section{Results of the calculation}
In Fig.~\ref{fig:chiral_potentials} we present the results of the calculation of the 
dielectric permittivity of graphene $\epsilon(r)$ which is the ratio of the bare potential (\ref{eq:potentials_large}) 
to the one measured on the lattice (\ref{eq:V_qq_def}). The dielectric permittivity is presented as a function of distance $r/a$ for 
the temperatures $T = 0.167,\,0.333,\,0.417,\,0.5\,$eV. Similar plots can be shown 
for the other temperatures under consideration.

It is seen from Fig.~\ref{fig:chiral_potentials} that $\epsilon(r)$ rises with the 
distance. Moreover the larger the temperature the sharper the rise of the dielectric permittivity. 
We believe that this behavior can be attributed to the Debye screening in graphene at nonzero temperature.
In order to get rid of the Debye screening effect and find the dielectric permittivity 
at zero temperature we are going to fit dielectric permittivity at every fixed distance 
with the anzatz 
\begin{equation}
	\label{eq:fit_temperature}
	\varepsilon(r, T) = A(r) + B(r) T + C(r) T^2.
\end{equation}
In conducting the fitting procedure we impose a constrain $B(r)>0$. Physically 
this constrain is motivated by the requirement that at $T\to0$ and $T\neq0$ Debye screening 
effect diminishes the potential i.e. enhances the value of the dielectric permittivity. 
The coefficient $A(r)$ in the last equation gives the dielectric permittivity of suspended graphene 
at zero temperature $\varepsilon(r, T = 0)$. In Fig.~\ref{fig:zero_potentials} and Tab.~\ref{tab:zero_potentials} we present the 
$\varepsilon(r, T = 0)$ as a function of distance $r/a$. 

From Fig.~\ref{fig:zero_potentials} and Tab.~\ref{tab:zero_potentials} it is seen that the dielectric permittivity 
at $r=0$ is $\epsilon=2.24\pm0.02$. Then it 
rises and after $r/a\ge3$ the dielectric permittivity reaches the plateau $\epsilon(r)\simeq4$. 
It is seen that the uncertainties of the calculation rise with the distance from rather small values  
to large ones. The uncertainties at distances $r/a\geq6$ become very large for this reason we do not show these points in Fig.~\ref{fig:zero_potentials}.

The main reason of large uncertainties of the calculation at large distance 
is the Debye screening at nonzero temperature in graphene. In order to find the value of the 
dielectric permittivity at large distances with better accuracy we have to account the Debye screening effect. 

We are going to do this as follows. The Debye screening of the Coulomb potential in graphene
was calculated in \cite{Braguta:2013rna, Astrakhantsev:2015cla} and it is given by
\beq
V(r)=\frac Q {\tilde \epsilon r} \int_0^{\infty} d \xi \frac {e^{-(m_D r) \xi}} {(1+\xi^2)^{3/2}},
\label{debye}
\eeq
where $\tilde \epsilon$ is the dielectric permittivity and the $m_D$ is the Debye mass. We fit our data for each temperature under consideration with
the parameters $\tilde \epsilon$ and $m_D$. Thus we get rid of the Debye screening effect 
which enhances the uncertainty at large distance. Notice that our bare potential deviates from the Coulomb (\ref{eq:potentials_large}) and tends to it only at large distance. 

In the fitting procedure we study the potential $V(r)$ in the region $r/a \in [4,8]$. 
We chose this region for the following reasons. Firstly, if one extends the region where 
our data is fitted to larger distance we will get larger uncertainties of the calculation.   
Secondly, within this region the deviation from the Coulomb potential is already sufficiently small $\sim 10-20\%$. 
The formula (\ref{debye}) describes our data quite well ($\chi^2/dof \sim 1$) for all temperatures and allows one to determine 
$\tilde \epsilon(T)$ as a function of temperature. 

To proceed we fit the results for the $\tilde \epsilon$ by the function $\tilde \epsilon(T) = \tilde A + \tilde B T + \tilde C T^2$.
The value of the $\tilde A$ gives the dielectric permittivity at zero temperature and large distance.
Thus we obtain
\beq
\epsilon=4.20 \pm 0.66 
\label{eps}
\eeq  
In the formula (\ref{eps}) we accounted statistical uncertainty and the uncertainty due to the deviation of the
potential (\ref{eq:potentials_large}) from the Coulomb.

Further let us proceed to the comparison of the results obtained in this paper with 
the perturbative expressions for the dielectric permittivity of suspended graphene. 
In Fig.~\ref{fig:zero_potentials} we present the one-loop calculation results of the dielectric permittivity. The red diamond points 
correspond to the one-loop dielectric permittivity calculated within the tight binding 
model (\ref{tbHam}) in \cite{Astrakhantsev:2015cla}. The blue line corresponds 
to the one-loop result (\ref{eps_one_loop}) which is obtained within the effective theory of graphene. 
It is seen that the full account of many body effects in the dielectric permittivity of suspended graphene within 
Monte Carlo study gives $\epsilon(r)$ which is very close to the one-loop results. 
The value of $\epsilon(r)$ at large distance (\ref{eps}) also agrees with one-loop result (\ref{eps_one_loop}). 

In Fig.~\ref{fig:zero_potentials} we also plot the results of \cite{fogler} 
which is given by the two-loop formula (\ref{eps_two_loop}). Since the calculation of $\epsilon$ in \cite{fogler}
was carried at two-loops, one has to renormalize the effective coupling constant $\alpha_{eff}$ at one-loop in order to get the value of the dielectric permittivity.
It is known that the renormalization of $\alpha_{eff}$ is reduced to the renormalization of the 
Fermi velocity $v_R$. It is rather difficult to find unambiguous expression for $v_R$, since the renormalized Fermi velocity depends on the infrared scale and 
in the problem under consideration a lot of scales can play a role of the infrared scale. 

To estimate $\epsilon$ at two loops we use the one-loop formula for $v_R$ obtained 
in \cite{Astrakhantsev:2015cla} and use temperature as the infrared scale
\beq
v_F^R=v_F \biggl [ 
1+ \frac 1 4 \frac {\alpha} {(v_F/c)} \log \biggr ( \frac {v_F \Lambda} {c T}   \biggl ),
\biggr ]
\eeq
where $v_F$ is the bare Fermi velocity, $c$ is the speed of light, $\Lambda$ is the ultraviolet cut-off 
and $T$ is the temperature which plays the role of the infrared scale in our estimation.  
In the calculation we take $v_F/c \sim 1/300$, $\Lambda\sim \hbar c/a$ and $T=0.1\,$eV which is a typical scale at which the 
calculations of this paper are carried out. With these numerical parameters we get $\epsilon=2.5$. 
If we carry out the calculation at the room temperature $T=293\,$K, the two-loop  
dielectric permittivity is $\epsilon=2.2$. 

One can estimate the two-loop result for $\epsilon$ as it was proposed in paper \cite{fogler}.
The authors of this paper used the momentum $q\sim \hbar/r$ as an infrared scale in the renormalization 
of the Fermi velocity. It is clear that in the limit $r\to \infty$, $\alpha_{eff} \to 0$ and $\epsilon \to 1$.
Notice that this limit is reached very slowly and one needs a very large graphene lattice to see that $\epsilon \simeq 1$.
For this reason one can ask what is the typical two-loop dielectric permittivity on the lattice which 
is used in our calculation. To estimate it we use the typical distance on the lattices under consideration
$L \sim 30 a \sim \hbar /q$. In this case we get $\epsilon=2.7$.

From this consideration one can state that all our estimations of the two-loop dielectric permittivity 
disagree with results obtained within Monte Carlo method. 
So, one can expect large higher order corrections to the two-loop result of \cite{fogler}.

\section{Conclusion}

In this paper the interaction potential between static charges in suspended graphene was studied
within the quantum Monte Carlo simulations. This approach is based on the tight-binding Hamiltonian 
without the expansion near the Fermi points what allows to avoid ambiguity due to the regularization procedure. 
In addition, the interactions between quasiparticles are parameterized by the realistic phenomenological 
potential, which deviates from the Coulomb potential at small distances. 
The main advantage of the Monte Carlo approach is that it fully accounts interactions between 
quasiparticles based on the first principles.

Within the Monte Carlo simulations we calculated the dielectric permittivity of suspended graphene for a set of 
temperatures. We carried out extrapolation to zero temperature for each distance between charges. 
Thus we calculated the dielectric permittivity of suspended graphene at zero temperature. 

We found that the behavior of the dielectric permittivity is the following. At zero distance the 
$\epsilon=2.24\pm0.02$. Then it rises and after $r/a\ge3$ the dielectric permittivity reaches the plateau $\epsilon(r)\simeq4.20\pm0.66$. The results obtained in this paper  
allow one to draw a conclusion that the full account of many body effects in the dielectric 
permittivity of suspended graphene gives $\epsilon(r)$ very close to 
the one-loop result obtained analytically within the tight-binding model of graphene. 
The value of $\epsilon(r)$ at large distances also agrees with the one-loop calculation
done within the low-energy effective theory of graphene. 

We also found that two-loop prediction for the dielectric permittivity deviates from the results of this paper. 
For this reason one can expect large higher order corrections to the two-loop prediction for the dielectric permittivity of suspended
graphene.

Finally would like to stress the fact that the full account of many-body effects in the dielectric 
permittivity of suspended graphene gives $\epsilon(r)$ close to the one-loop result
is highly nontrivial. The point is that the interaction in graphene is strong and it cannot 
be accounted by perturbation theory. For instance, if we substitute the interaction 
potential (\ref{eq:potentials_large}) by Coulomb with the same $A$ for all distances, 
graphene will turn from the semimetal phase to the insulator phase \cite{Buividovich:2012nx}.
So, it is not quite clear why the two-loop and higher order corrections should not modify 
the dielectric permittivity considerably compared to the one-loop calculation. From this perspective the dielectric 
permittivity of graphene is similar to the optical conductivity in graphene,
where higher order many-body corrections are also insignificant \cite{Boyda:2016emg}.

\section*{Acknowledgements}

We would like to thank Oleg Pavlovsky who proposed us to use the bare potential in the from (\ref{eq:potentials_large}).
The work of MIK was supported by Act 211 Government of the Russian
Federation, Contract No. 02.A03.21.0006. The work of MVU was
supported by DFG grant BU 2626/2-1. The work of VVB and AYN, which consisted of
numerical simulation and calculation of the static potential,
was supported by grant from the Russian Science Foundation (project number 16-12-10059).
This work was carried out using computing resources of the federal collective usage center ``Complex for simulation and data processing for mega-science facilities'' at NRC "Kurchatov Institute", http://ckp.nrcki.ru/. In addition we used the supercomputer of 
Institute for Theoretical and Experimental Physics (ITEP).

\end{document}